\documentclass[
&preprint,
longbibliography,
amsmath,amssymb,
aps,
prb,
twocolumn
]{revtex4-2}
\usepackage{graphicx}% Include figure files
\usepackage{dcolumn}% Align table columns on decimal point
\usepackage{bm}% bold math
\usepackage{epsfig}
\usepackage{afterpage}
\usepackage{lineno}

\def\be{\begin{equation}}
\def\ee{\end{equation}}
\def\bea{\begin{eqnarray}}
\def\eea{\end{eqnarray}}
\def\ba{\begin{array}}
\def\ea{\end{array}}
\def\bdm{\begin{displaymath}}
\def\edm{\end{displaymath}}

\begin{document}

\title{Broken time reversal symmetry vestigial state for a two-component superconductor in two spatial dimensions}% Force line breaks with \\

\author{P. T. How}
\email{pthow@outlook.com}
\affiliation{Institute of Physics, Academia Sinica, Taipei 115, Taiwan}
\author{S. K. Yip}
\email{yip@phys.sinica.edu.tw}
\affiliation{Institute of Physics, Academia Sinica, Taipei 115, Taiwan}
\affiliation{Institute of Atomic and Molecular Sciences, Academia Sinica, Taipei 115, Taiwan}

\date{\today}% It is always \today, today,
% but any date may be explicitly specified

\begin{abstract}

We consider the vestigial phase with broken time-reversal symmetry above the superconducting transition temperature of a two-component superconductor in two spatial dimensions.
We show that, in contrast to 3D, a vestigial phase is in general allowed within Ginzburg-Landau theory.   The vestigial phase
occupies an increasing temperature region if the parameters in the Ginzburg-Landau theory gives a larger energy difference between
the broken time-reversal symmetry phase and the other ordered phase. 

%\begin{description}
%\end{description}

\end{abstract}

\maketitle

%\tableofcontents
%\linenumbers

%\twocolumn

\section{Introduction}{\label{Sec:Intro}

Consider a superconductor with order parameter $(\Phi_1, \Phi_2)$ belonging to a two-dimensional representation,
here $\Phi_{1,2}$ are complex fields which transform as $(\Phi_1', \Phi_2') = e^{ i \chi} (\Phi_1, \Phi_2)$ 
under gauge transformation by $\chi$, and among themselves under spatial operations, such as 
$(\Phi_1',\Phi_2') = ( \Phi_1 \cos( \phi) + \Phi_2 \sin(\phi), 
\Phi_2 \cos( \phi) - \Phi_1 \sin(\phi ) )$ under rotations by $\phi$,
where $\phi$'s are restricted to multiples of 
$\frac{ 2 \pi}{3}$ ($\frac{ 2 \pi}{6}$) for trigonal (hexagonal) systems.
Such mutli-dimensional order parameters have been considered extensively since superfluid $^3$He \cite{LeggettRMP},
heavy fermion superconductors \cite{Volovik,Sauls}, and also more recently in many other systems  \cite{Yonezawa,Ghosh}.
In the two-dimensional representation case, 
below the superconducting transition temperature, depending on details of the free energy,
the energy minimum can be achieved by having $(\Phi_1, \Phi_2) \propto (1, 0)$, $(0, 1)$, or
$(\Phi_1, \Phi_2) \propto ( 1, \pm i)$.    In the former case, the order parameter breaks
gauge invariance and rotational invariance, whereas in the latter, gauge invariance as well as time-reversal
invariance.  In each case,  as oppose to the case of an order parameter $\Phi$ belonging to a one-dimensional representation,
there is additional symmetry breaking other than the gauge symmetry.

At high temperatures we have a symmetric phase with no broken symmetries.  
Within mean-field theory, at the superconducting transition temperature, 
the system goes into a phase with more than one symmetry broken.  
In principle at least, in a more complex scenario such as when fluctuations are included,
these broken symmetries do  not have to occur at the same time.
%While the expectation values $<\Phi_{1,2}>$
%remains zero, (i.e. above the superconducting transition),  some symmetry-violating higher-order
%combination of $\Phi_{1,2}$ may acquire a non-zero expectation value.
In particular, one can have a phase where say the rotational  (or time-reversal) symmetry is
broken, whereas the gauge symmetry is still intact.  In this case, we have the expectation values
$\langle \Phi_{1,2} \rangle = 0$, whereas, e.g.,
$\langle \Phi_1^* \Phi_1 - \Phi_2^* \Phi_2 \rangle \ne 0$
(or $i \langle( \Phi_1^* \Phi_2  - \Phi_2^* \Phi_1) \rangle \ne 0$). 
In additional to the above scenarios, one can have the possibilities 
that some other symmetry-violating combinations of $\Phi_{1,2}$ acquiring non-zero expectation values. 
For example, we can have 
$\langle \Phi_1^{2} + \Phi_2^{2} \rangle \ne 0$ even though $\langle \Phi_{1,2} \rangle = 0$.
In this latter case, while the order parameter is not preserved under general gauge transformations,
it is preserved under special transformation $\chi \to \chi + \pi$,  thus describes ``4e" pairing
\cite{note6e}. 
Such phases, often called ``vestigial" phases or phase with
``composite" or ``higher-order" parameters,
are gaining attention in the recent literature
\cite{VReview,Hecker18,Cho,Grinenko21,Fernandes21,Jian21,Hecker23}, 
though they have been investigated already in the past in similar 
\cite{Ashhab,Natu,Radic,Fischer16}
and related 
(e.g. \cite{Babaev04,Kuklov04,Herland10,Bojesen13,Bojesen14})
context. 
Besides superconductivity, these exotic phases are also relevant to other, e.g., magnetic, systems \cite{Gauthe,Song23}. 

In a previous paper \cite{How23}, considering 
three spatial dimensions, we show that, witin a Ginzburg-Landau theory with thermal fluctuations,
  such a vestigial phase is in general not stable,
except for the case of extreme gradient energy terms in the free energy.   
This is because, when the temperature is lowered so that the completely symmetric phase
is no longer the free energy stable minimum, either
(i) no saddle point corresponding to the vestigial phase exists, so that one has
a direct second order phase transition to the superfluid phase with broken
rotational (time-reversal) symmetry as well as gauge symmetry, or
(ii)  the vestigial phase with composite order parameter
is only a saddle point but fails to be a free energy minimum.  Instead, the free energy minimum
occurs in the region where the expectation value of $ \Phi_{1,2}$ is/are finite.
The system thus makes a joint first order phase transition into the superconducting case.
A similar situation can be shown to occur for a multicomponent Bose gas \cite{He20,How24}.

In this paper, we consider instead two spatial dimensions.  We show that the situation
becomes quite different.  Case (i) above remains a possibility, but for case (ii),
the saddle point does become stable in general for a finite range in temperature,
due to a very different free energy landscape. 
%We further investigate in detail this vestigial phase.  
Similar strong dependence on the spatial dimensionality has also been found in e.g., Ref \cite{Bojesen13, Bojesen14}. 

We shall mainly be studying the broken time reversal symmetry 
state. Our calculations will be presented in Sec \ref{Sec:Th}.  In Sec \ref{Sec:Con}, we shall
include also a short discussion on the nematic case and the ``4e'' state as well as conclusions. 
 
\section{Theory for Vestigial Order}\label{Sec:Th} 

We consider the same Hamiltonian as in ref \cite{How23}, but only now in two spatial dimensions.
We employ the ``spin-$\frac{1}{2}$'' notation, thus
\be
\label{Phi}
{\bf \Phi} = \left( \begin{array}{c} 
\Phi_{\uparrow} \\ \Phi_{\downarrow} \end{array} \right)
\ee

The effective Hamiltonian density 

\begin{equation}
{\mathcal H} = {\mathcal H_K} + {\mathcal H_{int}}
\end{equation}
consists of two parts.  For the  ``kinetic" part, we shall simply take

\be
{\mathcal H_K} = \sum_{s = \uparrow, \downarrow} \left[ \alpha \Phi_{s}^* \Phi_{s} 
  + K \left( \frac{ \partial \Phi_{s}^*} {\partial x_i}  \frac{ \partial \Phi_{s}} {\partial x_i} \right) \right]
\label{HK}
\ee
thus ignoring possible more complicated spatial ($x_{1,2}$) derivatives. 
(We remind the readers that \cite{How23} established the absence of the vestigial phase in three
spatial dimensions when the gradient energy is of this form). 
 Here $\alpha = \alpha(T)$ is positive (negative) above (below) a mean-field transition temperature which
we shall label as $T_0$, thus $\alpha(T) \approx \alpha' (T - T_0)$ with $\alpha' > 0$. 
The interacting part reads
\begin{equation}\label{intpm}
{\mathcal H_{int}}  =  \frac{ g_1 }{2} (|\Phi_{\uparrow}|^4 + |\Phi_{\downarrow}|^4) + g_2 (|\Phi_{\uparrow}|^2 |\Phi_{\downarrow}|^2) 
\end{equation}
$\Phi_{\uparrow,\downarrow}$ are related to $\Phi_{1,2}$ 
in the Introduction via $ \Phi_{\uparrow,\downarrow} = \frac{1}{\sqrt{2}} [ \Phi_1 \pm i \Phi_2]$.  
See also  \cite{How23} and  \cite{convert}.

If we simply minimize ${\mathcal H}$, the system is in the completely symmetric (normal) phase if $\alpha > 0$,
and for $\alpha < 0$, we have the broken time-reversal symmetry state $\Phi_{\uparrow} \ne 0$, $\Phi_{\downarrow} = 0$, 
$ \vert \Phi_{\uparrow}\vert^2 = \frac{ \vert \alpha \vert}{g_1}$ (or $\uparrow \leftrightarrow \downarrow$). with free energy density
$ - \frac{ \alpha^2}{ 2 g_1}$.  The ``nematic phase", with $\vert \Phi_{\uparrow} \vert = \vert \Phi_{\downarrow} \vert$,
is also a local energy minimum but has free energy $ - \frac{\alpha^2} { (g_1 + g_2)}$.
The stability of these mean-field states require $g_1 > 0$, $g_2 > - g_1$, and we shall restrict ourselves to this region.  
We shall also mainly deal with $g_2 > g_1> 0$ in this paper, so that the broken time reversal symmetry state has
the lower energy. 

%If no thermal fluctuations need to be considered (e.g., at $T=0$),  we can simply minimize this effective Hamiltonian.  The ground state has $\Phi_s$ uniform, with
%$\Phi_{\uparrow} \ne 0$ and $\Phi_{\downarrow} = 0$ (or vice versa), which represents thus a state
%with broken time-reversal symmetry and long range order. 

At finite temperatures, we need to consider the partition function \cite{PP,Z}
$Z \equiv \int_{\Phi_s} e^{ - \int d^2 x {\mathcal H} / T } $, where $\int_{\Phi_s}$ means sum over all 
configurations of $\Phi_{s} (\vec r)$. 
We employ the Hartree-Fock (HF) approximation.  The effective Hamiltonian density becomes
\be
{\mathcal H_{eff}} =  {\mathcal H_K} - h_{\uparrow} \Phi_{\uparrow}^* \Phi_{\uparrow} - h_{\downarrow} \Phi_{\downarrow}^* \Phi_{\downarrow}
\label{H0}
\ee
where $h_{\uparrow, \downarrow}$ are the self-energies
(not to be confused with external magnetic fields),  which are to be obtained self-consistently.  
$h_{\uparrow} \ne h_{\downarrow}$ signals that $\langle  \Phi_{\uparrow}^* \Phi_{\uparrow} \rangle \ne \langle  \Phi_{\downarrow}^* \Phi_{\downarrow} \rangle $.
 $ h_{\uparrow} - h_{\downarrow}$ thus serves as an order parameter for the broken Z$_2$ symmetry. 

After Fourier transform, 
\begin{equation}\label{Hkpm}
{\mathcal H_{eff}} =  \sum_{s, \vec k} \Phi_{\vec k, s}^{*} \left( \alpha + K k^2 - h_s \right) \Phi_{\vec k, s}
\end{equation}
%where $s$ represents $\uparrow, \downarrow$ and 
where $\vec k$ represents the wavevector. 
We thus have the expectation values
\begin{equation}\label{Gdef}
\langle \Phi_{\vec k, s} \Phi^*_{\vec k, s'} \rangle  =   T G_{s} (\vec k) \delta_{s, s'}
\end{equation}
with the ``Green's function" 
\begin{equation}\label{G-1}
 G_{s} (\vec k) = \frac{ 1}{  \alpha + K k^2 - h_s  } 
\end{equation}
For the vestigial phase, we must have $\alpha - h_s > 0$. 

The free energy density is, within the HF approximation,
\begin{widetext} 
\bea
{\mathcal F} & = &  \frac{T}{L^2} \sum_{\vec k, s} \left[ \ln (\alpha + K k^2 - h_s)    + h_s G_s(\vec k) \right]  \nonumber  \\
& &  + g_1 \left[ \left(\frac{T}{L^2} \sum_{\vec k} G_{\uparrow} (\vec k) \right )^2 +  \left( \frac{T}{L^2} \sum_{\vec k} G_{\downarrow} (\vec k) \right) ^2 \right]
 + g_2  \left[ \left( \frac{T}{L^2} \sum_{\vec k} G_{\uparrow} (\vec k) \right) \times \left( \frac{T}{L^2} \sum_{\vec k} G_{\downarrow} (\vec k) \right) \right]
\eea

This expression is ultraviolet divergent, both because of the $\ln (\alpha + K k^2 - h_s)$ and interaction terms.  
These divergences are also present even for $F \equiv F_0$ where we set $h_s$ to be zero (and thus 
replace $G_s (\vec k) $ by $G_0 (\vec k)  \equiv \frac{1}{ \alpha + K k^2}$). 
However, we note that insertion of the Hartree-Fock self-energies $( 2 g_1 + g_2) \frac{T}{L^2} \sum_{\vec k}  G_0(\vec k)$
to the propagators $G_s$ or $G_0$ would amount to  
replacing $\alpha$  by $\alpha + ( 2 g_1 + g_2) \frac{T}{L^2} \sum_{\vec k}  G_0(\vec k)$, which
can be regarded as a redefinition of $\alpha$.  
Using this renormalized $\alpha(T)$, the difference of the free energy density between the phase under consideration and 
$F_0$ can then be written as  \cite{How23}
\bea
\Delta { \mathcal  F} 
 & = &  \frac{T}{L^2} \sum_{\vec k, s} \left[ \ln (\alpha + K k^2 - h_s)  - \ln (\alpha + K k^2 )    + h_s G_s (\vec k) \right]  \nonumber  \\
& &  + g_1 \left[ \left( \frac{T}{L^2} \sum_{\vec k} (  G_{\uparrow} (\vec k) - G_0 (\vec k))  \right )^2 +
  \left( \frac{T}{L^2} \sum_{\vec k} ( G_{\downarrow} (\vec k) - G_0(\vec k))  \right) ^2 \right] \\
& + & g_2  \left[ \left( \frac{T}{L^2} \sum_{\vec k} ( G_{\uparrow} (\vec k) - G_0 (\vec k) ) \right) \times \left( \frac{T}{L^2}  \sum_{\vec k} ( G_{\downarrow} (\vec k) - G_0 (\vec k)  ) \right) \right] \nonumber
\eea

 This expression is ultraviolet convergent, and the contributions giving rise to finite $\Delta {\mathcal F}$ arise only
for small wavevectors when $\alpha$ and $h_s$ are small, as it should be.  

The momentum sums can be easily evaluated. For 3D, we reproduce the result in \cite{How23}. In the present case, we get
\bea
\Delta { \mathcal  F }& =& -  \frac{T}{ 4 \pi K} \left\{ \left[ \alpha \ln ( 1 - \frac{h_{\uparrow}}{\alpha}) + h_{\uparrow}  \right] + 
         \left[ \alpha \ln ( 1 - \frac{h_{\downarrow}}{\alpha} ) + h_{\downarrow}  \right] \right\}  \nonumber \\
 & & + \frac{T^2 g_1}{ ( 4 \pi K )^2 } \left[ \left(  \ln ( 1 - \frac{h_{\uparrow}}{\alpha} ) \right)^2 + \left(  \ln ( 1 - \frac{h_{\downarrow}}{\alpha} )\right)^2 \right]   
   + \frac{T^2 g_2}{ ( 4 \pi K)^2 } \left[  \ln ( 1 - \frac{h_{\uparrow}}{\alpha} ) \  \ln ( 1 - \frac{h_{\downarrow}}{\alpha} ) \right] 
\label{DF}
\eea
\end{widetext}

An important point to note is that, in contrast to the three dimensional case \cite{How23}, this free energy diverges to $+ \infty$ due to the 
$g_1$ term (since $g_1 > 0$)
when $h_s \to \alpha_{-}$.  Hence there is no ``falling off" to the unphysical ($\alpha - h_s < 0$) region, in contrast to \cite{How23}, and
stable non-trivial minima can exist within the physical $h_s < \alpha$ region.  See Fig 1. 

%%%%%%%%%%%%%%%%%%%%
\begin{figure}[h]\label{fig1}
%\begin{figure}
\begin{center}
\includegraphics[width=0.9\columnwidth]{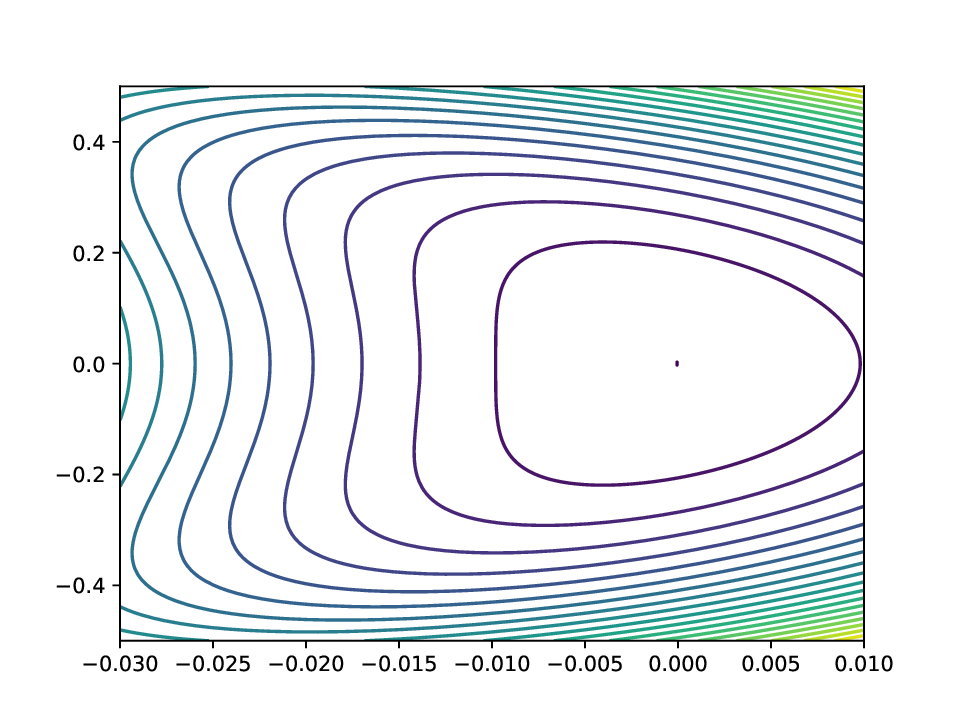}
\includegraphics[width=0.9\columnwidth]{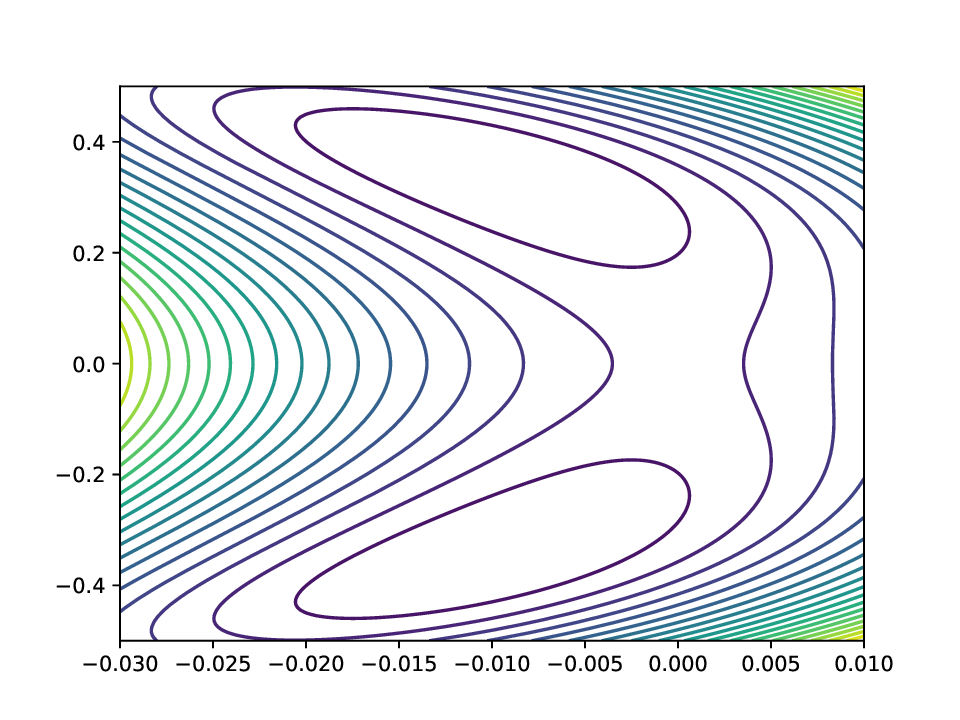}
\caption{
%(Color online)
Example contour plots of the free energy in eq (\ref{DF}). 
Upper diagram:  symmetric phase, lower:  broken symmetry phase. 
Abscissa:  $[\ln(1 - \frac{ h_{\uparrow}}{\alpha}) + \ln ( 1 - \frac{ h_{\downarrow}}{\alpha}) ] /2$,
ordinate:   $[\ln(1 - \frac{ h_{\uparrow}}{\alpha}) - \ln ( 1 - \frac{ h_{\downarrow}}{\alpha}) ] /2$. 
}
\label{fig1}
\end{center}
\end{figure}

Expansion of the free energy in terms of $h_z \equiv ( h_{\uparrow} - h_{\downarrow} )/2$ and
$h_0  \equiv ( h_{\uparrow} + h_{\downarrow} )/2$ gives
\be
\Delta {\mathcal F }=  a h_z^2 + b h_z^4 +   \gamma h_0 h_z^2 +  c h_0^2
\label{expand} 
\ee
where
\be
a = T I_2 [ 1  + T ( 2 g_1 - g_2) I_2 ]
\label{a}
\ee
\be
b = \frac{3}{2} T I_4 + 2 T^2 g_1 ( I_3^2 + 2 I_2 I_4) + T^2 g_2 ( I_3^2 - 2 I_2 I_4)
\ee
\be
\gamma = 4 T I_3 + 2 T^2 ( 6 g_1 - g_2) I_2 I_3
\ee
\be
c = T I_2 [ 1 + T (2 g_1 + g_2)  I_2 ] 
\ee
Here $I_2 = \frac{1}{ 4 \pi K \alpha}$ and generally
%$I_n  \equiv \frac{1}{L^2}  \sum_{\vec k} \frac{1}{ (\alpha + K k^2)^n } $ 
$I_n =\frac{1}{ ( n-1) 4 \pi K \alpha^{n-1}}$ for $n \ge 2$.

The coefficient $a$ changes sign at $T$ at $T_2$ where
\be
 0 = 1 + ( 2 g_1 - g_2) \frac{ T_2}{ 4 \pi K \alpha(T_2)} 
\label{T2}
\ee
 signalling a phase transition (at $T_2$ if second order).  This transition thus exists only when $g_2 - 2 g_1 > 0$. 
Eq (\ref{expand}) implies $h_0 = - \frac{\gamma}{2 c} h_z^2$. Eliminating $h_0$, the effective coefficient for $h_z^4$ 
becomes  $b - \frac{\gamma^2}{4 c}$.  The value of this coefficient  at $T_2$ is given  by
$b_{eff} =  \frac{T_2}{ 4 \pi K \alpha^3(T_2) } \frac{ 6 g_1 - g_2}{ 2 4 g_2}$ hence
positive only when $g_2 < 6 g_1$.  Hence transition is second order
only when  $g_2 < 6 g_1$ \cite{note3D}.  See Appendix for further analysis on this point. 
Below we shall confine ourselves only to this parameter regime.  
Since $\alpha$ is rapidly varying with temperature near $T_0$, eq (\ref{T2}) implies
\be
T_2 \approx T_0 \  [ 1  + \frac{ ( g_2  - 2 g_1)} { 4 \pi K \alpha'} ]
\label{T2e}
\ee
hence a transition temperature increasing  from $T_0$  linearly with $g_2 - 2 g_1$ when
the latter is positive.   Below T$_2$, $h_z^2 \approx - \frac{a' }{ 2 b_{eff}} ( T - T_2) $, with
$a' = - \frac{T_2}{4 \pi K \alpha^3(T_2)} \alpha'$.

The above has assumed that the transition is to a state with uniform $h_{0,z}$.  One can
also consider the free energy $F$ for the case where the self-energies $h_s$ varies with position.
If these fields have wavevector $\vec Q$, then the free energy has 
the form
\be
\Delta {\mathcal F }  =  a(Q)  h_z(\vec Q) h_z (- \vec Q) + ...
\label{FQ}
\ee
with
\be
a ( Q) = T I_2 (Q) [ 1  + T ( 2 g_1 - g_2) I_2 (Q) ]
\label{aQ}
\ee
where 
\be
I_2 (Q)  \equiv \frac{1}{L^2}  \sum_{\vec k} \frac{1}{ (\alpha + K k_+^2) ( \alpha  + K k_{-}^2 ) }
\label{I2Q}
\ee
with $\vec k_{\pm} = \vec k \pm \frac{\vec Q}{2}$. 
$I_2(Q) = I_2$ if $Q=0$, decreases with increasing $Q$ or $\alpha$, and is positive definite if $\alpha > 0$. 
Hence if $2 g_1 - g_2  > 0$, $a(Q)$ is positive for any $Q$ and positive $\alpha$.  
If  $2 g_1 - g_2  < 0$, $a(Q) > 0$ for all $Q$'s at high temperatures, and at $T_2$,
$a(Q)$ changes sign at $Q=0$ with $a(Q)> 0$ at $Q \ne 0$, verifying
that the transition is to the uniform state.

%More precisely,  $\Delta F$ itself should be considered as an effective Hamiltonian where $h_{0,z}$ are
%fluctuating quantities.  
 The above considerations show that,
for long wavelength fluctuations of $h_z$, the free energy density has the form
\be
\Delta {\mathcal F} =   a h_z^2 + \tilde K (\vec \nabla h_z)^2 +  b_{eff} h_z^4 
\label{tH}
\ee
The coefficient $\tilde K$ can be obtained from an expansion of $a(Q)$ at small $Q$. 
Using $I_2(Q) = I_2 - \alpha \frac{K Q^2}{2} I_4(0)$, 
we obtain thus
\be
\tilde K =  - \frac {\alpha K}{2} I_4  [ 1  + 2 T ( 2 g_1 - g_2) I_2 ]
\label{tK}
\ee
Near $a=0$ ($T_2$ given in eq (\ref{T2})), $\tilde K \approx \frac{T}{ 12 \pi \alpha} > 0$ \cite{signK}.
Eq (\ref{tH}) represents an effective Hamiltonian for a second order Ising transition,
with $\tilde K > 0$ and $b_{eff} > 0$ (the latter holds if $g_2 < 6 g_1$, as already mentioned).

The above considerations find the minimum of the free energy in $h_{z}$. More precisely, $h_z$ is 
itself a fluctuating quantity and 
the free energy density in eq (\ref{tH}) should be regarded as the effective Hamiltonian density for $h_z (\vec r)$.
We thus obtained an effective $\phi^4$ theory for the Ising transition where $h_z$ plays the role of the order parameter
for the $Z_2$ transition.
The considerations so far thus give, upon lowering of temperature,  an Ising transition
from a completely symmetric phase 
%(phase A in Fig 1)  
to a $Z_2$ broken symmetry phase
($h_z \ne 0$ yet with $\langle \Phi_s  \rangle= 0$)  at 
$T_2 > T_0$ (thus $\alpha > 0$) given by eq (\ref{T2e}) if $g_2 > 2 g_1$.
This is our vestigial phase. 
In this region, $h_{\uparrow} \ne h_{\downarrow}$, but both $\Phi_{\uparrow}$ and $\Phi_{\downarrow}$ have vanishing expectation values.
Correlations between $\Phi_s$ at different positions decay exponentially in space:
$\langle \Phi^*_{s}(\vec r) \Phi_s (\vec r') \rangle \propto e^{ - \frac{ \vert \vec r - \vec r' \vert }{ \lambda_s}}$.
Moreover, due to the finite $h_z$, $\lambda_{\uparrow} \ne \lambda_{\downarrow}$. 
Upon lowering of the temperature, $h_{z,0}$ both grows in magnitude, whereas $\alpha$ decreases. 
Within the above considerations, at temperature where $\alpha = h_{\uparrow}$, the
system makes a transition to the state with  $\langle \Phi_{\uparrow} \rangle \ne 0$ but $\langle \Phi_{\downarrow} \rangle = 0$ or vice versa,
a state just like $T=0$.   At this temperature, $\alpha - h_{\downarrow} > 0$ so that 
$\langle \Phi^*_{\downarrow}(\vec r) \Phi_{\downarrow}  (\vec r') \rangle$ still decays
exponentially 
%$ \propto e^{ - \frac{ \vert \vec r - \vec r' \vert }{ \lambda_{\downarrow}}}$. 
% This state is denoted by C in Fig 1. 
Furthermore, for $g_2 < 2 g_1$, $h_s$ vanishes, the system goes from the symmetric phase
to the state  $\langle \Phi_{\uparrow} \rangle \ne 0$ but $\langle \Phi_{\downarrow} \rangle = 0$ or vice versa
at $T_0$, where $\alpha $ vanishes  \cite{phaserule}.

At finite $T$, the phase with long range order 
%exist in Fig 1
just described is due to the artifact that phase fluctuation of $\Phi_s$ was not
considered.    Mermin-Wagner theorem states that this long range order is destroyed  in 2D.  
However, quasi-long range order \cite{KT} is allowed. 
For the phase diagram, the simplest possibility is that the
above mentioned phase with long range order is instead characterized by power law correlations, 
thus instead of finite expectation value for  $\Phi_{\uparrow}$,  we have
simply $\langle \Phi^*_{\uparrow}(\vec r) \Phi_{\uparrow} (\vec r') \rangle \propto \frac{1} {  \vert \vec r - \vec r' \vert ^{ \eta}}$.
%At least for large $g_2 - 2 g_1$, we expect that the Hartree-Fock gap  discussed before still exist, thus
%the correlation between $\Phi_{\downarrow}$ still decays exponentially with distance. 
The resulting phase diagram is as given in Fig 2a. 

Another possibility is that, due to thermal fluctuations of the phase, there is always a vestigial Z$_2$ broken symmetry phase
that lies between the completely symmetric phase and the quasi-long range order phase, 
even for the region $g_2 < 2 g_1$.   This possibility has been raised in a few theoretical calculations
based on models which are related to though not the same as the one we have in this paper \cite{Bojesen13,Zeng24,Liu24,Maccari23}
(though there are also related studies where such a phase is absent \cite{Song22}). 
%These investigations all correspond to the case where amplitude fluctuations are ignored and 
%only phase fluctuations are considered \cite{note}.  
 The resulting vestigial phase again only has short range order, but since Z$_2$ is broken, the decaying lengths
$\lambda_{\uparrow,\downarrow}$ are thus unequal.  This phase is indistinguishable
from our vestigial phase described by $h_{z} \ne 0$, though the physical picture
giving rise to this broken Z$_2$ symmetry seems quite different. 
The resulting phase diagram is sketched qualitatively in Fig 2b.  \cite{multi}
For both Fig 2a and Fig 2b, the phase transition temperatures all vanish
at $g_2 = g_1$.  At this point there the symmetry is enhanced to SO(3),
which forbids any order at finite temperature in two spatial dimensions,
a fact also pointed out in \cite{Maccari23}.

%%%%%%%%%%%%%%%%%%%%
\begin{figure}[h]\label{fig2}
%\begin{figure}
\begin{center}
\includegraphics[width=0.9\columnwidth]{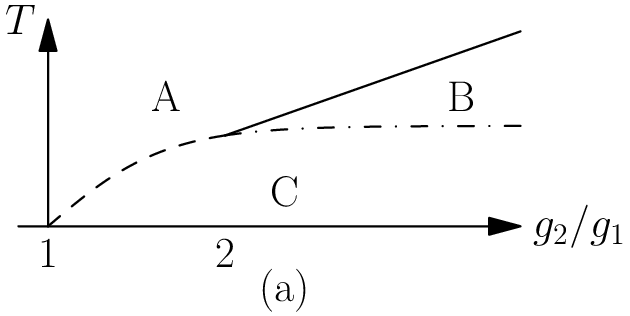}
\includegraphics[width=0.9\columnwidth]{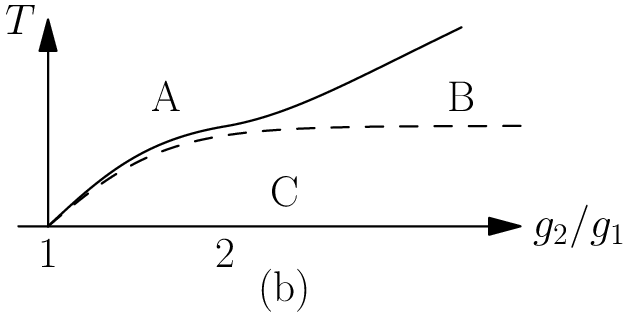}
\caption{
%(Color online)
Possible phase diagrams. Region A is the symmetric phase.  Region B is the vestigial phase with broken Z$_2$ symmetry,
but with only short range order for both $\Phi_{\uparrow}$ and $\Phi_{\downarrow}$. 
$\langle \Phi^*_{s}(\vec r) \Phi_s (\vec r') \rangle \propto e^{ - \frac{ \vert \vec r - \vec r' \vert }{ \lambda_s}}$.
but $\lambda_{\uparrow} \ne \lambda_{\downarrow}$. 
In Region C, one of the $\Phi_{s}$ has quasi-long-range order while the other has only short range correlation. 
Full lines:  Ising transitions, dashed or dot-dashed:  BKT transitions
}
\label{fig2}
\end{center}
%\end{figure}
\end{figure}
%%%%%%%%%%%%%%%%%%%%

\section{Conclusion}\label{Sec:Con}

   Starting from a Ginzburg-Landau theory for a two-component superconductor, we show that the vestigial broken time reversal symmetry state
wtih no superconducting order parameter is possible in two spatial dimensions, provided that
the parameters lies in the situable region.  This is in strong contrast to the case in 
three spatial dimensions \cite{How23}, where such as phase is in general not possible except
for some extreme situations. We also obtain the effective $\phi^4$ theory for
this Ising transition in terms of the parameters entering this Ginsburg-Landau theory.

  Similar calculations can be extended also to vestigial nematic order,
 governed by an order parameter  $\vec h = (h_x, h_y)$. 
We have already shown in \cite{How23} that the vestigial nematic state is generally unstable in 
three spatial dimensions.  Back to the present case of two spatial dimensions, 
calculations similar to Sec \ref{Sec:Th} can also be carried out. 
For example, we still have eq. (\ref{expand}) etc  if we exchange 
$h_z$ there by $ \vert \vec h \vert $, provided we also replace
%$h_{\uparrow, \downarrow}$ by $\pm \vert \vec h \vert + h_0$. 
$g_{1,2}$ there by $\frac{g_1 + g_2}{2}$ and $g_1$ respectively
({\it c.f.} also \cite{How23}),
%  For example, eq (\ref{a}) holds if we replace
hence $2 g_1 - g_2$ in eq (\ref{a}) by $g_2$.
A vestigial nematic state thus requires  $g_2 < 0$.
$b_{eff}$ is now proportional to $ \frac{2 g_1 + 3 g_2}{g_1}$.
The  effective gradient energy has the form $ \tilde K (\partial_i h_j) (\partial_i h_j)$ 
(note our eq (\ref{HK}) has no ``spin-orbit" coupling) with coefficient $\tilde K$ given  by the same as the expression  below eq (\ref{tK}). 
Instead of an Ising transition, we expect a Kosterlitz-Thouless transition for $\vec h$ itself
when $g_2 > - \frac{2}{3} g_1$, but a more complicated scenario is feasible if this inequality 
is not satisfied. 

If $g_2 < 0$, we can also have ``$4$e" superconductivity with ``pairing'' between fields 
 $\Phi_{\uparrow}$ and $ \Phi_{\downarrow}$.  
Vestigial ``4e'' state now corresponds to quasi-long range order of the product
$\Phi_{\uparrow}  \Phi_{\downarrow}$ ($ \propto \Phi_1^{2} + \Phi_2^{2} $) but without quasi-long range order of either $\Phi_s$. 
When the gradient term is simply taken as in (\ref{HK}), the calculations for the effective free energy is entirely parallel to 
that of the nematic phase, as has already been pointed out in \cite{Fernandes21,Jian21,Hecker23}. 
Discussion in the last paragraph also applies in this case with appropriate substitutions.

\section{Acknowledgements}

This work is supported by the Ministry
of Science and Technology, Taiwan under Grant No.  MOST-110-2112-M-001-051 -MY3, and P.T.H. is supported under
Grant No. MOST 112-2811-M-001-051.

% \tilde {\mathcal H}_{eff}

\vspace{0.5 cm}

\newcounter{seq}

\newenvironment{seq}{\refstepcounter{seq}\equation}{\tag{B\theseq}\endequation}

\appendix
%\begin{center}
%{\bf APPENDIX}\label{App}
%\end{center} 

\section{Order of phase transition}\label{App}

We analyze this phase transition  without expansion in $h_{0,z}$.
We define $x_{s} = - \ln ( 1 - \frac{h_s}{\alpha} )$, where
we have chosen the sign so that $x_s$ is an increasing function of $h_s$.
 All $h_s < \alpha$, hence $ - \infty < x_s < \infty$ are acceptable.
Employing  $x = ( x_{\uparrow} + x_{\downarrow})/2$,
and $y = ( x_{\uparrow} - x_{\downarrow})/2$, 
eq (\ref{DF}) can be written as 
\begin{widetext}
\be
\frac{ \Delta {\mathcal F }} {\alpha T / ( 4 \pi K) } = 2 \left[ x +   e ^{-x} \cosh(y) - 1 \right]
 + ( 2 \tilde g_1 + \tilde g_2) x^2 + ( 2 \tilde g_1  - \tilde g_2 ) y^2 
\ee
\end{widetext}
where $\tilde g_{1,2} = \frac{ T g_{1,2}}{ 4 \pi \alpha K}$. 

The stationary point conditions are 
\be
0 = ( 1  -  e^{-x} \cosh y )  + ( 2 \tilde g_1 + \tilde g_2 ) x
\ee
and 
\be
0 =  e^{-x} \sinh y    -  (  \tilde g_2 - 2 \tilde g_1  ) y
\ee
The second equation is trivially satisfied by $y=0$.  For $y \ne 0$, we solve for   $x$ using this second equation
and substitute back to the first to yield a single equation for $y$:
\be
G  = \frac
{\frac{y}{ \tanh y}  -  \frac{\alpha (T)}{\alpha(T_2) } }
{ \ln \left[ \frac{ \sinh y}{y}   \frac{\alpha (T)}{\alpha(T_2) } \right] }
\ee
with $G \equiv  \frac{  2 \tilde g_1 + \tilde g_2}{ \tilde g_2 - 2 \tilde g_1} $.
Since we have $ 0 < 2 \tilde g  <  \tilde g_2$, $G$ decreases with increasing  $ \tilde g_2/ \tilde g_1$. 
For $ 2 \tilde g_1  <  \tilde g_{2} < 6 \tilde g$, $G$ lies between $2$ and $+\infty$.
For  $ 6  \tilde g_1  <  \tilde g_{2}  $, $G$ lies between $1$ and $2$. 
Graphical solution  shows that for $2 < G < \infty$, $y$ vanishes for $\alpha  > \alpha(T_2)$.
A non-trivial solution for $y$ starts from zero and grows with decreasing $\alpha < \alpha(T_2)$,
thus a typical second order phase transition at $T=T_2$. 
  For $1 < G < 2$, finite $y$ solutions already exist at some 
$\alpha(T)  > \alpha(T_2)$, and with decreasing  $\alpha(T)$, one obtains two solutions, one with
$y$ decreasing and the other increasing with decrasing $\alpha$. $\alpha(T_2)$ is the point at which
the decreasing solution approaches $y=0$,   which
shows a typical first order transition behavior.  Note however in contrast to 3D, we have
 a local free energy minimum, not just a saddle point.


\begin{thebibliography}{50}
\bibitem{note6e}  More complex situations such as ``6e" etc pairing are possible, but we shall not go into those in this paper.
\bibitem{convert}  With $\Phi_{\uparrow,\downarrow} = \frac{1}{\sqrt{2}} (\Phi_1 \pm i \Phi_2)$, 
then ${\mathcal K} = \left[ \alpha \Phi_{\nu}^* \Phi_{\nu} +  
K  \left( \frac{ \partial \Phi_{\nu}^*} {\partial x_i}  \frac{ \partial \Phi_{\nu}} {\partial x_i} \right) \right]$
while 
${\mathcal H_{int}} = \frac{\beta_1}{2} (\Phi^*_{\mu} \Phi_{\mu} ) (\Phi^*_{\nu} \Phi_{\nu} ) 
+ \frac{\beta_2}{2} (\Phi^*_{\mu} \Phi^*_{\mu} ) (\Phi_{\nu} \Phi_{\nu} ) $
with implicit sum for  $\mu$ and $\nu$ over $1$ and $2$,
and $\beta_1 = g_1$ and $\beta_2 = ( g_2 - g_1) /2$.   In \cite{How23}, this was taken to be the model
of a superconductor with order parameter $(\Phi_1, \Phi_2)$ belonging  to a two-dimensional representation. 
This model can also be viewed as describing two superconductors $\Phi_1$ and $\Phi_2$, with
the $\beta_2$ term contains a contribution representing a pair tunneling term
 (as well as others).
\bibitem{note3D} The same calculation for 3D, with generally $I_n \equiv \frac{1}{L^d}  \sum_{\vec k} \frac{1}{ (\alpha + K k^2)^n } $ and $d=3$ instead of $2$,
 shows that $b_{eff}$ is always negative 
(given by $ -  \frac{T}{ 8 \pi \alpha^{5/2} K^{3/2}} \frac{ (g_2 - 2 g_1)}{ 32 g_2}$ ).   This result corroborates that in 
\cite{How23}, showing that we can only have a joint first order phase transition.
\bibitem{signK} It can be seen from eq (\ref{tK}) that $\tilde K$ is negative if $2 g_1 > g_2$,
and also at sufficiently high temperatures if $g_2 > 2 g_1$.  However, as explained already
below eq (\ref{I2Q}), this does not lead to transitions to non-uniform states.
\bibitem{phaserule}  This phase diagram cannot be quite right, even within mean-field theory for the $\Phi_{s}$ transitions.
It predicts three second order phase transitions line meeting at a point with finite angles with respect to each other, 
and the finite $\Phi_{\uparrow}$ phase occupies a region
with angle larger than $\pi$, violating the phase rules in \cite{Yip} and \cite{Leggett} respectively. 
We however would not pursue this question further since the 2D nature requires us to consider phase fluctuations
of $\Phi_s$, to be discussed in the next paragraph.
\bibitem{multi} In the scenarios of \cite{Zeng24} and \cite{Liu24}, the transition from the symmetric phase to
the Z$_2$ broken phase is via a BKT transition, instead of an Ising transition we deduced for $g_2 > 2 g_1$.
In that case, for Fig 2b, there must also be an additional multicritical point separating the two regions.
\end{thebibliography}
\end{document}